\documentclass[manuscript,screen, nonacm]{acmart}

\usepackage{datetime}
\usepackage{xspace}
\usepackage{booktabs}
\usepackage{multirow}
\usepackage{subcaption}
\usepackage{bbm}

\newcommand{\HTTPLong}{{\textsf{HTTP-9Years}}\xspace }
\newcommand{\HTTPLast}{{\textsf{HTTP-Dec23}}\xspace }

\newcommand{\accepted}{{\textit{Accepted-All}}\xspace }
\newcommand{\partials}{{\textit{Custom}}\xspace }
\newcommand{\rejected}{{\textit{Rejected-All}}\xspace }

\newcommand{\acceptbutton}{{\texttt{Accept~All}}\xspace }
\newcommand{\custombutton}{{\texttt{Custom Permissions}}\xspace }
\newcommand{\rejectbutton}{{\texttt{Reject~All}}\xspace }

\begin{document}

\title[Privacy Policies and Consent Management Platforms]{Privacy Policies and Consent Management Platforms:\\
Growth and Users' Interactions over Time}

\author{Nikhil Jha}
\email{nikhil.jha@polito.it}
\orcid{0000-0002-7533-4051}
\affiliation{%
  \institution{Politecnico di Torino}
%  \streetaddress{C.so Duca degli Abruzzi, 24}
%  \city{Torino}
  \country{Italy}
%  \postcode{10129}
}

\author{Martino Trevisan}
\email{martino.trevisan@dia.units.it}
\orcid{0000-0002-4258-4679}
\affiliation{%
  \institution{Università degli Studi di Trieste}
%  \streetaddress{Via Alfonso Valerio, 6/1}
%  \city{Trieste}
  \country{Italy}
%  \postcode{34127}
}

\author{Marco Mellia}
\email{marco.mellia@polito.it}
\orcid{0000-0003-1859-6693}
\affiliation{%
  \institution{Politecnico di Torino}
%  \streetaddress{C.so Duca degli Abruzzi, 24}
%  \city{Torino}
  \country{Italy}
%  \postcode{10129}
}

\author{Daniel Fernandez}
\email{daniel@illow.io}
\author{Rodrigo Irarrazaval}
\email{rodrigo@illow.io}
\affiliation{%
  \institution{illow.io}
%  \streetaddress{360 NW 27th St}
%  \city{Miami}
%  \state{Florida}
  \country{United States}
%  \postcode{33127}
}

\renewcommand{\shortauthors}{Jha, et al.}

\begin{abstract}
In response to growing concerns about user privacy, legislators have introduced new regulations and laws such as the General Data Protection Regulation (GDPR) and the California Consumer Privacy Act (CCPA) that force websites to obtain user consent before activating personal data collection, fundamental to providing targeted advertising. The cornerstone of this consent-seeking process involves the use of Privacy Banners, the technical mechanism to collect users' approval for data collection practices. Consent management platforms (CMPs) have emerged as practical solutions to make it easier for website administrators to properly manage consent in ever-changing scenarios, allowing them to outsource the complexities of managing user consent and activating advertising features.

This paper presents a detailed and longitudinal analysis of the evolution of CMPs spanning nine years. We take a twofold perspective: Firstly, thanks to the HTTP Archive dataset, we provide insights into the growth, market share, and geographical spread of CMPs. Noteworthy observations include the substantial impact of GDPR on the proliferation of CMPs in Europe, where more than 40\% of websites currently adopt a CMP. Secondly, we analyse millions of user interactions with a medium-sized CMP present in thousands of websites worldwide. We observe how even small changes in the design of Privacy Banners have a critical impact on the user's giving or denying their consent to data collection. For instance, over 60\% of users do not consent when offered a simple ``one-click reject-all'' option. Conversely, when opting out requires more than one click, about 90\% of users prefer to simply give their consent. The main objective is in fact to eliminate the annoying privacy banner rather the make an informed decision. Curiously, we observe iOS users exhibit a higher tendency to accept cookies compared to Android users, possibly indicating greater confidence in the privacy offered by Apple devices. We believe the findings of this paper contribute to a deeper understanding of the multifaceted interactions between privacy regulations, technological solutions and user choices in the evolving Web ecosystem. We also show that the availability of large open datasets, although not explicitly designed and collected for our goals, is fundamental to exploring different angles of the internet evolution over time.\footnote{Part of this work previously appeared at the 2023 Network Traffic Measurement and Analysis Conference (TMA).}.
\end{abstract}


\maketitle

\section{Introduction}

The landscape of Web advertising has been continuously evolving since the Web is more and more intertwined with people's lives, and many websites base their revenues on advertisements (ads). To maximize ad effectiveness, Interest-Based Advertisement (IBA) stems from the fundamental assumption that users will be more interested in ads that relate to their specific interests. In the Web ecosystem, this fueled the deployment of mechanisms that aim at building specific per-user profiles to push the IBA to a great level of detail. Third-party cookies have largely been at the center of the IBA ecosystem. Once installed on the end user's browser, the so-called \emph{profiling cookies} cookies enable a third party to collect information on the visits a user performs on those websites where the third-party is present~\cite{mayer2012third,metwalley2015online,englehardt2016online}. This allows the third party to collect information about the users' navigation history and derive what they are interested in. 

Over the years, the use of cookies raised concerns for the users' privacy, and legislators started acting to regulate the abuse of these practices across the Web to increase the users' awareness of the process. In the European Union, the first significant step in this direction was the introduction of the ``Cookie Law''~\cite{directive2009} in 2009, which states that every website using first- or third-party cookies should obtain user's approval via a \textit{Privacy Banner}. The measure was further enforced in 2018 with the General Data Protection Regulation (GDPR)~\cite{gdpr}, which introduced severe fines to non-compliant websites and organizations. Similarly, other legislators around the globe started regulating the collection of personal data, making the presence of Privacy Banners pervasive as we all experience while surfing the Web. Notably,  California's California Consumer Privacy Act (CCPA)~\cite{ccpa} in 2018, the Brazilian General Personal Data Protection Law (LGPD)~\cite{lgpd} and the Canadian Personal Information Protection and Electronic Documents Act (PIPEDA)~\cite{pipeda} in 2000 (and currently being reformed~\cite{billc27}).

To avoid the burden of creating, controlling, and checking the compliance of Privacy Banners with existing and evolving regulations, websites often outsource the banner creation and management to external companies offering \textit{Consent Management Platforms} (CMPs). CMPs take care of the technical complexity of implementing Privacy Banners, offering their customers a simple mechanism to install and manage the user's consent and data collection at large.

In this paper, we present a longitudinal measurement study that aims to understand how the Consent Management Platforms landscape has evolved over the years, and how users interact with them. We consider two angles in our study: the system and the end-user point of view.
From the system point of view, we break down the global evolution of CMPs, observing the growth of different solutions, their market share, their effectiveness in controlling the usage of third-party cookies and their spread in different geographical regions. For this, we leverage the unique opportunity offered by the HTTP Archive~\cite{httparchive} which includes the history of millions of websites since 2015.

From the end-user perspective, we leverage the preferences expressed by users when interacting with a medium-sized CMP, observing the impact of different options like the presence (or absence) of a ``one-click-choice'' option, the position of the banner, the device they use, the country they are from, etc. To this end, we leverage a unique dataset of the illow.io\footnote{\url{https://illow.io}} CMP which spans more than two years of data since its introduction in the CMP market, collecting more than 20 million interactions with the CMP. 

Our study unveils the introduction of the GDPR and similar regulations fuelled the explosion of the CMPs which are now present in 40\% of websites in Europe. We also witness an increase in websites that opted to directly remove third-party tracking technologies, probably balancing the benefit and complexity of managing their presence. Surprisingly, more than 50\% of websites adopting a CMP still install third-party cookies of possible tracking services before the user accepts their usage. This is in fact due to the complexity of managing third-party content, and the lack of standard technical solutions to enforce the user's choice by CMPs. Moreover, a website must nowadays adapt its behavior with third parties depending on regulations and the country of the user. This, in turn, calls for more analysis and for more extensive archival projects.

From the user's point, we observe that more than 60\% of users deny the usage of third-party tracking cookies when offered a simple ``one-click reject-all'' action. Yet, the large majority of these choices derive from the intent of quickly removing the intrusive banner from the screen rather than an explicit and conscious decision. In fact, in those countries where the opt-out choice requires more than one click, about 90\% of users accept the usage of cookies via a one-click accept-all button. Curiously, iOS users tend to accept the usage of cookies more frequently than Android users. This may be linked to a higher confidence in privacy offered by iOS and Apple devices.

Both the HTTP Archive and the illow.io datasets have not been collected nor been designed specifically for this work. Yet, their availability allows us to study and present interesting insights in an opportunistic manner. In particular, HTTP Archive collects its data uniquely from locations in the U.S., which is nowadays limiting, as the web is more and more customized based on users' location to adapt to the different privacy regulations. To allow the reproducibility of part of our results, we publish online the pre-processed HTTP Archive dataset and the code to obtain the figures of this paper~\cite{opendata}.

The remainder of the paper is organized as follows. In Section~\ref{sec:related}, we introduce the regulation history and discuss related works. In Section~\ref{sec:metho}, we discuss the methodology and the datasets we use for our analysis. In Section~\ref{sec:cmps-in-the-web} we present how the pervasiveness of the CMPs in the Web has evolved over time while in Section~\ref{sec:user-interaction} we focus on  users' interactions when facing the Privacy Banners of the illow.io CMP. Finally, in Section~\ref{sec:conclusion}, we summarize the main finding of our work.
\section{Privacy Regulations and Related Works}
\label{sec:related}

Behavioural advertising has always been a pillar of the Web ecosystem and entailed the massive collection of personal information through web tracking and cookies~\cite{metwalley2015online, pujol2015annoyed, englehardt2016online, acar2014web, rizzo2021unveiling, papadogiannakis2021, sipior2011online, mayer2012third, estrada2017online}. The implications of web tracking on users' privacy encouraged the legislator to issue privacy-related regulations. 

In the European Union, the introduction of the Cookie Law~\cite{directive2009} in 2009 and updated in 2013 led to the proliferation of Privacy Banners~\cite{eijk2019impact}. The regulation states that when visiting a website, users have to interact with the Privacy Banner to explicitly opt-in to the usage of tracking mechanisms. Only after getting explicit users' consent, the website (and any third party embedded in the website) are allowed to install cookies and start the data collection. Privacy Banners, however, do not fully protect users in many cases~\cite{trevisan20194}. Later in 2018, the GDPR profoundly influenced the Internet user experience~\cite{sanchezrola2019can, dabrowski2019measuring, linden2020privacy, degeling2019we, kretschmer2021cookie} for EU-based users by defining severe sanctions for violators.\footnote{U.K. adopted GDPR in the ``Data Protection Act'' in 2018. In the rest of the paper, we refer to ``GDPR countries'' as any European country where the GDPR is in place, including U.K.}

Other countries issued similar regulations, with notable examples in the Brazilian Lei Geral de Proteção de Dados Pessoais (LGPD)~\cite{lgpd} (entered into force on September 18\textsuperscript{th} 2020), 
the PIPEDA~\cite{pipeda} (April 13\textsuperscript{th} 2000) and Quebec 25 law~\cite{law25} (September 22\textsuperscript{nd} 2021) in Canada,
and the California Consumer Privacy Act (CCPA)~\cite{ccpa} (January 1st 2020) which spurred other US states to enforce similar regulations.
At the high level, these regulations provide individuals control over the personal data that businesses collect about them. In a nutshell, they mandate that users have to explicitly opt-in to the collection of personal data, including the usage of cookies and other tracking technologies on the web.

A large number of websites started to rely on Consent Management Platforms, i.e., external companies that provide technical solutions to manage users' consent collection by offering customizable and easy-to-deploy Privacy Banners. Hils \emph{et al.}~\cite{hils20measuring} provided a first analysis of CMPs' popularity. By actively crawling popular websites for almost three-years they provide a detailed picture of the growth in popularity of CMPs,  observing websites switching CMPs, and manually analyzing the privacy policies they publish to check the purpose for data collection indicated. Our work complements and extends this analysis by offering a longitudinal analysis over a period of 9 years thank to the HTTP Archive, and by including the users' perspective on the analysis.

In general, it has been shown that most users tend to ignore privacy-related notices~\cite{vila2003we, grossklags2007empirical, coventry2016personality}, up to getting annoyed by these. This behavior has gone under the name of ``privacy paradox'': Users claim to be concerned about their privacy, while at the same time taking little actions to protect their data~\cite{barth2017privacy}. 
In fact, given the advertising-based business model, some websites and CMPs make efforts to increase the cookie acceptance rate. Recent works have shown that banners often nudge users to acceptance by exploiting dark patterns in the user interface, if not openly disregarding GDPR's requirements~\cite{matte2020do}, or making it difficult for users to exercise their rights~\cite{habib2020scavenger}. Some hide the  content of a website, which is visible only upon cookie acceptance. These impacts automated internet measurements since the first visit to a website may not show the actual website content~\cite{jha2022internet}.
Nudging includes offering the user a \acceptbutton default button via intrusive banners~\cite{CookieBenchmarkStudy,bauer2021you}, which is often the case~\cite{hausner2021dark} with websites presenting large pop-ups or wall-style banners that cover most of the webpage content. Researchers have shown that apparently minor design choices have a significant effect on inducing the user to accept the cookies~\cite{soe2020circumventing, nowes2020dark, vanbavel2016testing, utz2019uninformed, kulyk2022so, singh2022cookie}. For instance, Habib \emph{et al.}~\cite{habib2022okay} compared the behaviour of $1\,109$ volunteers on 12 different banners, showing that the cookie acceptance rate varies by $\approx 25\%$ depending on banner design. Unequal paths (i.e., \emph{refuse} action more complicated than the \emph{accept} action) and blocking banners are the most important factors. Our work complements this by analyzing the behaviour of millions of users when facing the banner of a popular CMP whose  banner assumes various forms across time and country. This allows us to compare the impact of the banner design including many more factors than those considered by Habib \emph{et al.}~\cite{habib2022okay}. 

The part of the paper focused on the study of user interaction the CMP is an extension and a refinement of a previous work~\cite{jha2023i}. In this work, we extend the timeline of our analysis to include more than two years. This allows us to better appreciate the impact of the banner design on the user choice and to include several new angles, including the impact of the user's device, country, etc.
\section{Methodology and datasets}
\label{sec:metho}

\begin{figure}
    \centering
    \includegraphics[width=0.8\columnwidth]{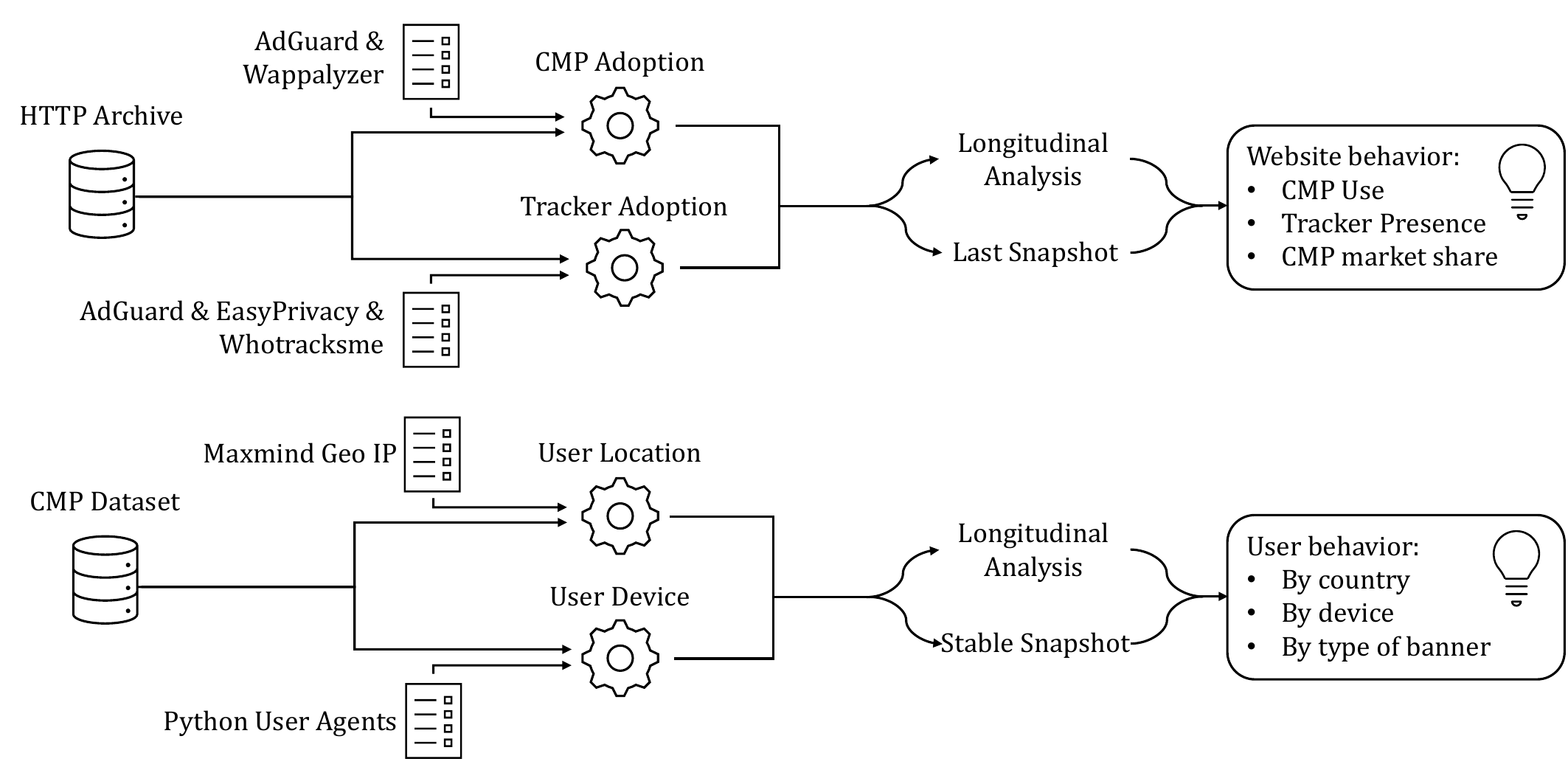}
    \caption{Data processing and analysis workflow. The top part refers to the CMP adoption study. The bottom parts to the user behaviour study.}
    \label{fig:processing_metho}
\end{figure}

This section describes our datasets and the processing methodology we adopt to carry out our analyses. We rely on two datasets to study CMP adoption and user behaviour on CMPs that we enrich using external sources. We provide a high-level overview of our data processing and analysis methodology in Figure ~\ref{fig:processing_metho}. The top part refers to the CMP adoption study which we base on the HTTP Archive. The bottom part refers to the user behaviour study which we base on the CMP dataset. In both cases, we start from a longitudinal dataset, and integrate it with external information to observe both the evolution over time of different metrics and the detailed breakdown in the most recent period. In the appendix, we provide the exact structure of the two datasets. We release the pre-processed HTTP Archive data used in our analyses along with the code to replicate our results as open-source~\cite{opendata}.

\subsection{CMP Adoption}

\subsubsection{Dataset}
~\newline
To study CMP adoption through the years we use the HTTP Archive open dataset~\cite{httparchive}. The curators of this archive visit a list of worldwide websites using an automated version of the Google Chrome browser. They perform the visits using test agents from various Google Cloud data centers in the U.S. The test agents use the WebPageTest\footnote{\url{https://webpagetest.org/} toolset, Accessed on \today.} to instrument Google Chrome and collect several metrics on the web page visit using the HTTP Archive (HAR) format, a JSON-formatted archive format for logging a web browser's interaction with a webpage. All websites are visited every 15 days to create a longitudinal archive. The curators update the list of websites they visit on regular periods. In detail, until June 2018 the website list contains the top-500~k entries of the Alexa rank. In July 2018, they extended the list to include the top-1M websites taken from the Chrome UX Report, which contains a variable set of entries ($\approx$ 10~M), updated on a monthly basis.\footnote{\url{https://developer.chrome.com/docs/crux}, Accessed on \today.}. The HTTP Archive data is available on Google Cloud from which we downloaded the data into a Spark cluster for processing. 

In this work, we rely on the monthly snapshots offered by HTTP Archive starting from January 2015 until December 2023. For the longitudinal study, we focus on the subset of websites that are present for the whole time. There are $47,899$ websites. We refer to this dataset as \HTTPLong. To study the impact of the countries the website belongs to, we use the top-level domain (e.g., \texttt{.fr} ) to associate a website with its main country. \footnote{Although there exist other approaches to link a website to its country (e.g., inspecting the registrar of the DNS domain or location of the server), using the country-code TLDs represents an accurate yet conservative option, even if we may erroneously exclude websites using global TLDs (\texttt{.com}, \texttt{.org}, etc.).} We have $11,819$ website whose top-level domain belongs to a GDPR-regulated countries, and that are present for all the nine yars. We focus most of our analysis on this set of websites.  Table~\ref{tab:httparchive} details the breakdown of sites for each country.

To complement the longitudinal study and detail the most recent picture of the CMP adoption, we resort to the last available snapshot which includes the top-1M websites as visited during December 2023. This dataset includes websites from all over the world. We refer to this dataset as \HTTPLast (last line in Table~\ref{tab:httparchive}).

For each visit, the HTTP Archive datasets report several statistics and details on each HTTP transaction carried out to fetch the webpage's objects. In detail, for each HTTP request/response pair, the dataset reports the URL, the timings and various HTTP headers including Cookies, Referrer, etc. This allows us to study the presence of various third-party elements in a website, including the presence of CMPs or other third-party services. To this end, we examine all HTTP requests issued during the page loading and extract the domains from the corresponding URL. In case the HTTP request refers to a third-party server (i.e., the domain is different from the webpage one), we match the domain into the lists described in the next paragraph to detect whether the website adopts a CMP or embeds a potential web tracker.

\begin{table}[]

\caption{Breakdown of Websites present in the HTTP Archive Dataset.}
\label{tab:httparchive}

\renewcommand{\arraystretch}{0.7}
\setlength{\tabcolsep}{15pt}

\begin{tabular}{@{}crlrc@{}}
\toprule
\multicolumn{3}{c}{\textbf{Dataset}}                                                                               & \textbf{Websites}  & \multicolumn{1}{l}{\textbf{Period}}                             \\ \midrule \midrule
\multirow{13}{*}{\rotatebox[origin=c]{90}{\HTTPLong}} & \textbf{Total}       &                                     & \textbf{47,899} & \multirow{13}{*}{\rotatebox[origin=c]{90}{Jan 2015 - Dec 2023}} \\ \cmidrule(lr){2-4}
                                                      & \textbf{GDPR}        &                                     & \textbf{11,819} &                                                                 \\
                                                      & \multicolumn{1}{l}{} & \multicolumn{1}{r}{\texttt{.de}}    & 2,907           &                                                                 \\
                                                      & \multicolumn{1}{l}{} & \multicolumn{1}{r}{\texttt{.es}}    & 603              &                                                                 \\
                                                      & \multicolumn{1}{l}{} & \multicolumn{1}{r}{\texttt{.fr}}    & 1,060           &                                                                 \\
                                                      & \multicolumn{1}{l}{} & \multicolumn{1}{r}{\texttt{.it}}    & 1,060           &                                                                 \\
                                                      & \multicolumn{1}{l}{} & \multicolumn{1}{r}{\texttt{.co.uk}} & 1,617           &                                                                 \\ \cmidrule(lr){2-4}
                                                      & EU No-GDPR           &                                     & 1,412           &                                                                 \\
                                                      & Africa               &                                     & 336              &                                                                 \\
                                                      & Asia                 &                                     & 3,926           &                                                                 \\
                                                      & North America        &                                     & 777              &                                                                 \\
                                                      & South America        &                                     & 1,399           &                                                                 \\
                                                      & Other                &                                     & 28,230          &                                                                 \\ \midrule \midrule
\multicolumn{2}{r}{\HTTPLast}                                                &                                     & 1,000,000      & \multicolumn{1}{l}{Dec 2023}                                    \\ \bottomrule
\end{tabular}
\end{table}

\subsubsection{Processing Methodology}

\paragraph{CMP Identification}

We study the adoption of CMPs by looking for URL requests to the most popular CMPs. For this, we build a  list of CMP domains by combining two sources: the  AdGuard~\cite{adguard} list which includes a set of CMP domain names; And the Wappalyzer~\cite{wappalyzer} list, a profiler tool that is used to identify various characteristics of a website including the presence of Content Management Platforms and Analytics products. Wappalyzer explicitly lists domain names of CMPs. We merge the two sets of CMP domains, and, given the small number of entries, we manually verify each one. In total, we obtain 84 CMPs, each identified by one or more domains.

\paragraph{Potential Trackers}
To detect the presence of potential third-party trackers in a given website, we merge publicly-available lists provided by Whotracksme~\cite{whotracksme} (an anti-tracking-related open-data provider), EasyPrivacy~\cite{easyprivacy} (one of the lists at the core of AdBlock tracker-blocking strategy) and AdGuard~\cite{adguard} (a popular ad-blocking tool). Trackers are identified by their domain name. For robustness, we merge the three lists and consider as a potential tracker any third-party domain that appears in at least two lists. In total, we obtain 1,497 domains that we consider tracking services.\footnote{For simplicity, we match the domains using the \emph{second-level domain name} --- i.e., a hostname truncated after the second label. We handle the case of two-label country-code TLDs such as \texttt{.co.uk}.}

We record the presence of a tracker during a visit if the webpage embeds an object served by a tracking domain and the latter installs a persistent cookie. The HTTP archive dataset does not indicate the content and the lifespan of the cookie, making it impossible to verify if the third-party cookie is actually a \emph{Profiling Cookie} or not. We thus verify the \emph{potential} presence of a Tracker. In fact, such a cookie could be a technical cookie used to store, e.g., consent information, not violating the privacy policy or the regulations.

\subsection{User interaction}
\label{sec:illow-dataset-method}

\subsubsection{Dataset}
~\newline
To study the users' interaction patterns when facing a Privacy Banner we use a proprietary dataset of illow.io, a medium-sized CMP company. The illow.io CMP provides web developers with the ability to include a simple Privacy Banner to control their preferences about data collection, via cookies or other means. The banner is shown the first time a user accesses the website as a small overlay window that can be placed in different locations on the screen. The users can express their preference across four categories of cookies: i) necessary, ii) statistical, iii) preferential, and iv) marketing cookies. The necessary cookies include those technical cookies that are needed for the site to function, including the storage of the user's preference on the usage of cookies, and the user cannot opt out from their usage. Once the users have expressed their choice, the CMP sets the necessary cookies on the user's device to store their preferences. When the user accesses again the website, no banner is shown and preferences can be updated via a dedicated page.

The Privacy Banner is served directly by the illow.io servers in the form of a small set of Java Scripts. When users interact with the Privacy Banner (i.e., provide their choice, or change their consent), the illow.io servers log the event. The collection happens only when users submit their preferences (as detailed in the privacy notice). No data is collected if the user does not perform any action on the banner. The logged information includes the time of the visit, a random ID of the entry, an anonymized ID for the website, the type and position of the banner, which cookies the user accepted, the \texttt{User Agent} header value as set by the browser, and the client IPv4 /24 subnet. We use the client's subnet to map a user to country and state via the MaxMind GeoIP\footnote{\url{https://www.maxmind.com/}, Accessed on \today.} database.\footnote{We do not consider IPv6, as it generates negligible traffic.} This information is necessary to implement the functionalities of the platform (i.e., record user's preferences for the next visits to the website), and it is useful to customize the information provided to users (e.g., show the banner in different language, decide what version of the banner to show to the user --- more details in Section~\ref{sec:illow-method}), and to collect statistics about the usage of the platform, including the billing of the website deploying the CMP. All these pieces of information, including the description of the collected data and the purpose of the collection, are documented in the privacy policy the CMP offers to users.

Users submitting (or changing) their preference generate an entry in the log, which we call an \textit{interaction} in the following. As previously stated, each entry is associated with a random ID, and, thus it is impossible to re-identify or track a user across different websites, guaranteeing users' privacy. To further protect the CMP customers, in the data used for this study, the website name is also anonymized by replacing the domain name with a random identifier.

We classify each interaction according to the combination of cookie categories accepted by the user. Specifically, we classify interactions as:

\begin{itemize}
    \item \accepted: if all cookie categories are accepted, either with a single click on the \acceptbutton button, by individually accepting all the cookies after clicking on \custombutton button, or by clicking on the \textit{Don't sell my data} link;
    \item \rejected: if only the necessary cookies are accepted, either by clicking \rejectbutton button (or equivalent action) if present, or by manually deactivating all the cookies after clicking on \custombutton (recall that necessary cookies cannot be excluded);
    \item \partials: if at least one among the statistical, preferential and marketing cookies is accepted through the \custombutton screen.
\end{itemize}

In the following, we compare the \rejected rate against different factors, such as banner layout and position, user's country and type of device. This offers us precious insight into understanding which banner layout is the most effective in allowing the users to exercise their right to refusal.

\begin{table}[]
    \caption{Number of users' interactions per geographical region - illow.io dataset.}
    \label{tab:volumes}
    \centering
    \normalsize
    \setlength{\tabcolsep}{15pt}
    \begin{tabular}{lrr}
        \toprule
        \bf Region      & \bf Total & \bf July-August '23 \\
        \midrule
        South America   &   11,953,295 (48.20\%) & 3,415,701 (43.79\%)\\
        GDPR-regulated  &    9,417,253 (37.97\%) & 3,124,085 (40.06\%)\\
        North America   &     2,448,467 (9.87\%) &   849,008 (10.89\%)\\
        Asia            &       377,853 (1.52\%) &   184,390 (2.36\%)\\
        Rest of Europe  &       367,107 (1.48\%) &   140,969 (1.81\%)\\
        Africa          &       149,905 (0.60\%) &    45,925 (0.59\%)\\
        Oceania         &        87,685 (0.35\%) &    38,757 (0.50\%)\\
        \bottomrule
    \end{tabular}
\end{table}

The dataset contains interactions spanning from July 1\textsuperscript{st}, 2022 to August 22\textsuperscript{nd}, 2023. In this period, the CMP collected $\approx25M$ interactions from more than 6,000 websites. We report the breakdown of interactions per user's regions in Table~\ref{tab:volumes}.
Also for this dataset, we consider the whole data when focusing on the evolution over time, while we consider the period from July 5\textsuperscript{th} 2023 until the end to consider a stable period and minimize the effect of the transients.

\paragraph{Banner Layouts}

\begin{figure}
  \begin{subfigure}{.33\textwidth}
    \centering
    \includegraphics[width=.9\linewidth]{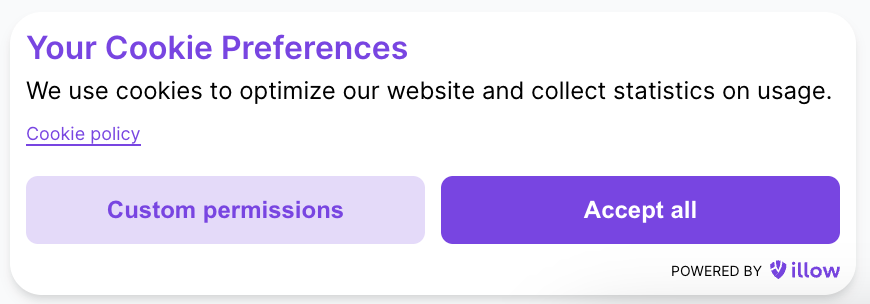}
    \vspace{2mm}
    \caption{General banner.}
    \label{fig:banner-general}
  \end{subfigure}  
  \begin{subfigure}{.33\textwidth}
    \centering
    \includegraphics[width=.9\linewidth]{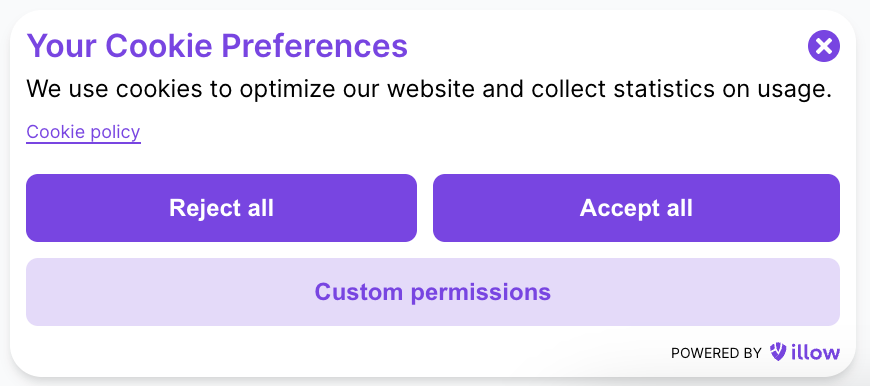}
    \caption{GDPR-compliant banner.}
    \label{fig:banner-gdpr}
  \end{subfigure}  
  \begin{subfigure}{.33\textwidth}
    \centering
    \includegraphics[width=.9\linewidth]{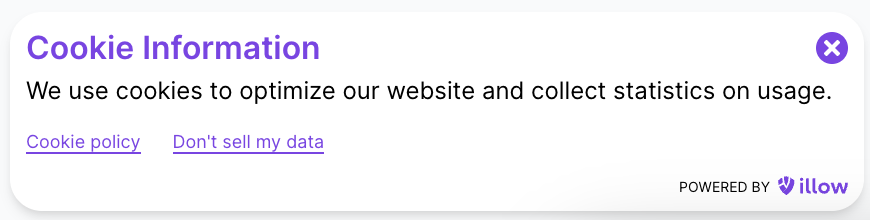}
    \vspace{4mm}
    \caption{CCPA-compliant banner.}
    \label{fig:banner-ccpa}
  \end{subfigure}  
  \caption{The available banners introduced since July 2023 in the illow.io CMP. Figures showcase the pop-up style banners.}
  \label{fig:banners}
\end{figure}

The illow.io CMP complies with the privacy regulations of different countries and follows the best practices recommended by the privacy regulation bodies. The different regulations specify how online companies should collect data of the visitors (from areas under their jurisdiction) and indicate how the user's consent should be obtained. When a new user visits a website, the CMP serves the correct type of banner according to the country and region the customer visit comes from.

Over the years, the illow.io CMP updated several times the design and actions offered by the banner to enable new functionalities or to adapt to evolving regulations. In our longitudinal study, we show the impact of such changes. The description below refers to the most recent configuration of the CMP which has been in place since July 5\textsuperscript{th} 2023. The illow.io CMP supports three different types of banners that comply with different regulations:
\begin{itemize}
    \item The \textit{General} banner (Figure~\ref{fig:banner-general}) presents two buttons, the \acceptbutton and the \custombutton. The first button gives the permission to install any type of cookie -- i.e., represent the opt-in choice. The second button brings to a selection window, where the user can fine-select the categories of cookies to accept. This is the default banner shown to users who connect from areas that are not covered by any particular regulation.
    \item The \textit{GDPR-compliant} banner (Figure~\ref{fig:banner-gdpr}) presents three buttons. The \acceptbutton gives the permission to install any type of cookie. The \rejectbutton only forbids all but necessary cookies. Finally, the \custombutton prompts to the selection window as in the \textit{General} banner. The banner also features a close-window button in the form of an ``X''. 
    The CMP offers this banner to users connecting from GDPR-regulated countries and from countries with similar regulations -- i.e., Canada (due to the PIPEDA and Quebec 25 laws) and Brazil (due to the LGPD). This banner was introduced in August 2022, while, before, users in these countries faced the \textit{General} banner. Moreover, the operation of the ``X'' close button changed from ``accept-all'' to ``reject-all'' action in December 2022. We will explore these changes in Section~\ref{sec:user-interaction}.
    \item The \textit{CCPA-compliant} banner (Figure~\ref{fig:banner-ccpa}) notifies the user that cookies are being used. Two links appear: the first brings to the website privacy policy; The second, named ``Don't sell my data'', brings the user to the selection window where the user can fine-select the active cookies. This banner is served by default to visitors coming from California and Utah, under the regulation of the CCPA and the Utah Consumer Privacy Act (UCPA), respectively. The banner offers a close-window button in the form of an ``X'' that closes the banner without requiring further action from the user, which implicitly accepts the use of cookies. Notice that the operation of this banner is different from the previous two: The user must explicitly opt out of cookies, which are enabled by default.
\end{itemize}

\subsubsection{Processing Methodology}
\label{sec:illow-method}

The dataset is securely stored on a big data cluster located in our institution. We process it using Python code to extract various metrics and statistics.
%We rely on the Python User Agent library to parse the HTTP user agent exposed by the user browser.
Given that an imbalance exists in the website audience, we want to prevent popular websites from biasing the results. For this, we opt to show results using a website-wise macro-average of the metrics under study. In other words, we compute the desired metric average for each website, separately. Then we compute the average over the websites. In such a way, each website weights one, regardless of the number of interactions it generates. We consider a per-website average valid if we observe at least 10 interactions from it, in at least 50 websites.

Formally, given a target metric $M$ (e.g., \rejected), a set of websites $w\in\mathcal{W}$, each with a population of interactions $\mathcal{I}_w$, an indicator function $\mathbbm{1}_M(i)$ which return 1 if $i$ refers to $M$, 0 otherwise (e.g., whether interaction $i$ records a \rejected choice or not), we define as $\overline{M}(\mathcal{I})$ the website-wise macro-average of $M$ computed over the samples belonging to the subset $\mathcal{I} = \bigcup_{w \in W: \lvert I_W \rvert \ge 10} \mathcal{I}_w$. Formally:
\begin{equation}
    \overline{M}(\mathcal{I}_w) =  \frac{1}{|\mathcal{I}_w|}\sum_{i \in \mathcal{I}_w} \left(\mathbbm{1}_M(i)\right), \text{ given } \lvert I_W \rvert \ge 10,
    \label{eq1}
\end{equation}

\begin{equation}
    \overline{M}(\mathcal{I}) = \frac{1}{|\mathcal{W}|} \sum_{w \in \mathcal{W}} \overline{M}(\mathcal{I}_w), \text{ given } \lvert \mathcal{W} \rvert \ge 50,
    \label{eq2}
\end{equation}
Equation(\ref{eq1}) computes the per website average. Equation(\ref{eq2}) computes the overall macro average.

In addition, we also evaluate the confidence interval of the macro average. Hence, each estimate is presented as:
$$\overline{M}(\mathcal{I}) \pm c \cdot \frac{\overline{S}\left( \mathcal{I} \right)}{\lvert W \rvert},$$
where $c$ corresponds to the quantile of a Student's $t$-distribution with $\lvert W \rvert-1$ degrees of freedom, and $\overline{S}\left( \mathcal{I} \right)$ is the sample standard deviation of each website-wise average. In this work, we consider a confidence interval of 90\% and report the confidence interval as an error bar. As our main target metric, we consider the \rejected rate.
\section{Temporal evolution of CMP adoption}
\label{sec:cmps-in-the-web}

In this section, we study the adoption of CMPs on websites during the last 9 years. To this end, we rely on the two HTTP Archive datasets (\HTTPLong and \HTTPLast) as described in Section~\ref{sec:metho}.

\begin{figure}
\centering
\begin{minipage}{.5\textwidth}
  \centering
\includegraphics[width=\linewidth]{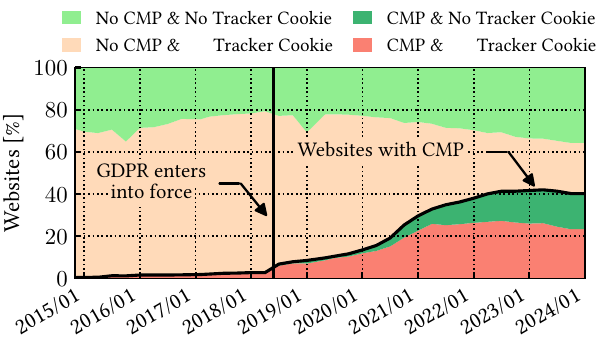}  \captionof{figure}{Fraction of websites with/without a CMP and contacting potential trackers. \HTTPLong Dataset of 11,819 websites that are present during the whole period in GDPR-regulated countries. }
  \label{fig:httparchive_websites_cmp_tracker}
\end{minipage}%
\qquad
\begin{minipage}{.45\textwidth}
    \centering
    \captionof{table}{Fraction of websites with/without a CMP and contacting potential trackers. \HTTPLast Dataset with top-1M website worldwide. }
    \label{tab:httparchive_global}
    \vspace{2mm}
    \small
    \begin{tabular}{l|cc|l}
    \textbf{}        & \textbf{Tracker} & \textbf{No Tracker} & \textbf{Overall} \\ \hline
    \textbf{CMP}     & \ \ 9.9\% (75.6\%)            & \ \ 3.2\% (24.4\%)               & 13.1\%           \\
    \textbf{No CMP}  & 40.2\% (46.2\%)           & 46.7\% (58.8\%)              & 86.9\%           \\ \hline
    \textbf{Overall} & 50.1\%           & 49.9\%              &                 
    \end{tabular}
\end{minipage}
\end{figure}

\begin{table}[]
    \centering
    \small
    \caption{Potential tracker and Third Party prevalence on websites adopting different CMPs. \HTTPLast Dataset with top-1M website worldwide.}
\begin{tabular}{lccr}
\toprule
        \bf CMP &  \bf Websites With Trackers [\%] &  \bf Average Trackers & \bf Websites \\
\midrule
   Usercentrics &                            32.94 &                  2.14 &    16,077 \\
        Iubenda &                            56.30 &                  3.05 &    13,378 \\
       OneTrust &                            61.33 &                  9.23 &    46,144 \\
   CookieScript &                            63.55 &                  4.78 &     2,093 \\
      Cookiebot &                            75.85 &                  4.99 &    12,415 \\
         Didomi &                            82.14 &                 20.29 &     5,202 \\
Funding Choices &                            97.85 &                 16.30 &    26,921 \\
        HubSpot &                            99.20 &                  9.07 &    20,871 \\
\bottomrule
\end{tabular}
    \label{tab:cmp_tracker}
\end{table}

\subsection{Overall trend}

We start our analysis with Figure~\ref{fig:httparchive_websites_cmp_tracker} where we show the fraction of websites with and without a CMP on the 11,819 European-based websites where GDPR is in force. We further break down the statistics to consider sites that embed or not a potential tracker. Observe how the CMP adoption (indicated by the black curve) was negligible before the GDPR (entered into force in May 2018, indicated with a black vertical line in the Figure): At that time, only $\approx$~5\% of websites adopted a CMP. Recall that some form of Privacy Banners was already implemented to comply with the previous  ``Cookie Law'' European directive. Yet, CMPs were not a common solution. Notice that between 2015 and 2018, at least 3 out of 4 websites used to embed at least a potential tracker even if the user did not accept their usage. This is in line with our previous finding~\cite{trevisan20194}. In fact, this figure is a lower bound for the popularity of tracking services because we expect more trackers would be contacted upon the user's consent (recall that the HTTP Archive test agents perform always a ``first-visit'' to all websites and do not interact in any form with any banner).

Starting from May 2018, the CMP adoption quickly increases, reaching 30\% at the beginning of 2021 and 42\% in late 2023. As expected, the CMP adoption erodes primarily the fraction of websites with no CMP and potential tracker (light red area) which shrinks from $\approx$70\% (May 2019) to $\approx$20\% (Dec. 2023). This indicates that the introduction of the GDPR created a new need for those websites using ads and tracking services to adopt solid privacy management solutions (i.e., a CMP).

It is interesting to notice that the fraction of websites without a CMP \emph{and} without potential trackers (light green area) increases as well, growing from 20\% (May 2019) to 35\% of websites (Dec. 2023). We argue this is due to two phenomenon: first, some large and popular websites started implementing proprietary consent management solutions (which we cannot detect). Second, smaller web services started balancing the benefit of including third-party ad services with the burden of correctly managing their presence, opting out from including trackers and deploying a CMP.

Focus now on the fraction of websites with CMPs (bottom solid red and green areas). Unexpectedly, more than 60\% of websites that adopt a CMP  still lets the user's browser install cookies from potential trackers before collecting any user's consent (bottom dark red area). We argue this is due to the complexity of managing advertisement and tracking platforms, coupled with the burden of installing and operating a CMP and the lack of a standard mechanism for CMPs to  control third-party services. In a nutshell, as previously found~\cite{trevisan20194}, the presence of a consent banner does not guarantee the correct management of data collection policies. Recall that we cannot claim those websites violate the GPDR: first, the HTTP Archive visits websites from the U.S. and a website/CMP might differentiate their behaviour when the user visit comes from a non-GDPR country. Second, a tracking service can legitimately install a technical cookie before the user's consent. As previously said, we cannot distinguish this case from the installation of an actual profiling cookie due to limitations of the HTTP Archive data.

We complement our analysis by looking at the most recent snapshot which contains the top 1 million ranked websites. Here, we broad the picture to include websites from all over the world, regardless of their country and of being present in the past snapshots. In Table~\ref{tab:httparchive_global} we show the CMP adoption and potential tracker share. Globally. Only 13.1\% of websites adopts a CMP, while 50.1\% of them  include potential trackers. As noticed before, most websites adopting a CMP embed potential trackers (9.8\% of the total, corresponding to 75\% of sites with CMP).

We investigate in more detail the surprisingly large fraction of websites with CMP that include a potential tracker. For the top-10 most popular CMPs (see later), in Table~\ref{tab:cmp_tracker}, we report the percentage of websites that embed a potential tracker (sorted from lowest to highest), their average number and the number of websites they are present in. Results are very heterogeneous, with some CMPs able to prevent the browser from contacting potential trackers more efficiently than others. For instance, 32.9\% of  websites adopting Usercentric still contact two potential trackers on average. Conversely, 58.8\% of sites adopting the Sourcepoint CMP still allow the visitors to contact more than 33 potential trackers. Again, due to the limitations of the HTTP Archive, we cannot confirm the cookies installed by potential trackers are profiling cookies. Recalling that the HTTP Archive visits come from the USA, we manually verified 10 websites adopting the Usercentric and 10 the Sourcepoint CMP, visiting them from Italy. In all cases, we observe both  CMPs correctly permit the browser to contact trackers only after accepting the site privacy policy. This call for further investigations and automated means to provide an extensive evaluation of the eventual violation of the privacy policies as this would depend on multiple factors such as client country location, language, device type, etc. as presented in our previous work~\cite{jha2022internet}. We leave this as future work.

\begin{figure}
\centering
\begin{minipage}{.45\textwidth}
  \centering
  \includegraphics[width=\linewidth]{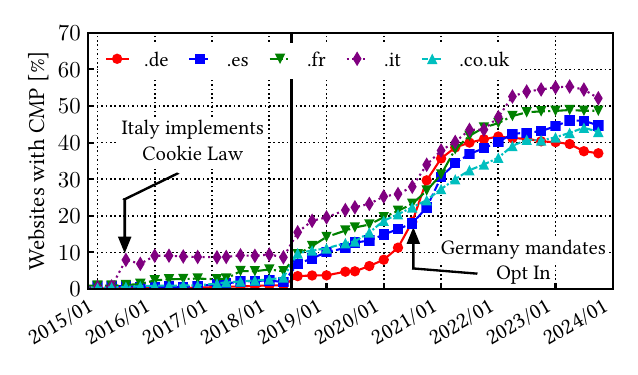}
  \captionof{figure}{Fraction of websites with a CMP for 5 European TLDs. \HTTPLong Dataset of 11,819 websites that are present during the whole period, in GDPR-regulated countries.}
  \label{fig:httparchive_websites_with_cmp_tld}
\end{minipage}%
\qquad
\begin{minipage}{.45\textwidth}
  \centering
  \includegraphics[width=\linewidth]{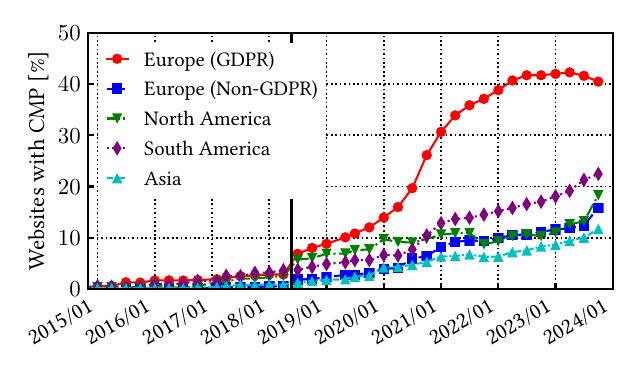}
  \captionof{figure}{Fraction of websites with a CMP for different continents. \HTTPLong Dataset of websites that are present during the whole period, in not GDPR-regulated countries.}
  \label{fig:httparchive_websites_with_cmp_continent}
\end{minipage}
\end{figure}

\subsection{Country-wise and region-wise diversity}

Different countries and regions of the world have adopted different regulations at different times. To observe how this impacts the evolution of CMPs, in Figure~\ref{fig:httparchive_websites_with_cmp_tld} we break down the CMP adoption for GDPR countries. Here, we focus on websites belonging to five country code top-level domains, namely: \texttt{.de}, \texttt{.es}, \texttt{.fr}, \texttt{it} and \texttt{.co.uk}, i.e., websites based in Germany, Spain, France, Italy and the United Kingdom, respectively. Before GDPR (represented by the black vertical line), CMP adoption was negligible in all countries except for Italy, where approximately 10\% of websites already adopted a CMP starting from mid-2015. This is due to the  implementation of the old European Directive called Cookie Law~\cite{directive2009}, which entered into force in June 2015 in Italy. Indeed, the Italian Data Protection Authority ``Garante per la protezione dei dati personali'' mandated the correct implementation of a Privacy Banner~\cite{notagarante} by  June 2, 2015. As a consequence, Italian websites resorted to CMPs to solve the issue. The Italian CMP Iubenda grabbed a market share of 90.3\% in Italy by the end of 2017. In the last part of this section, we analyse in detail the market share of popular CMPs. 

Following the entering into force of the GPDR (May 2018), the CMP adoption started increasing gradually until 2023, even if at a different pace. Italian websites (purple line) grow quicker, remaining those with the highest rate until the most recent snapshot. Conversely, the German websites (red line) have the lowest CMP adoption rate reaching just $\approx$ 10\% in mid-2020. From June 2020, they exhibit a very rapid increase, reaching 35\% by the end of the year. This acceleration is due to the decision of the German Federal Court~\cite{notadsk} that in May 2020 modified the GDPR in a more restrictive sense. The judgement underlined the need for \textit{explicit} user consent before installing marketing or nonessential cookies and specified that an opt-out mechanism is invalid. We speculate this led to the massive adoption of CMPs in Germany, causing their sudden growth to 40\% by mid-2021.
Spain, France, and the UK exhibit a more gradual increase. The CMP adoption was below 10\% in May 2019 and increased to 40-50\% by the end of the observation period.

Interestingly, we observe that, in all countries but the UK, CMP adoption is no longer increasing in 2023, and we even notice a moderate yet measurable decrease in Italy and Germany.

For the sake of comparison, in Figure~\ref{fig:httparchive_websites_with_cmp_continent} we dissect CMP adoption for different regions of the world. The red solid line refers to GDPR countries (i.e., it is equivalent to the back curve of Figure~\ref{fig:httparchive_websites_cmp_tracker}). In these countries, CMP adoption tops 42\%, while we observe a moderately decreasing trend in the second semester of 2023. In the remaining European countries (those not implementing the GDPR, blue line), the CMP adoption rate stands at significantly lower values, and even with an increasing trend, it has not yet reached 20\%. We observe similar numbers for North America and Asia -- green and cyan curves, respectively. Interestingly, in South America, the CMP adoption rate is consistently higher over time, exceeding 20\%  since mid-2023. This is mostly due to Brazilian (\texttt{.br}) websites, which represent 59\% of the 1,399 South American websites in the \HTTPLong dataset. Recall that in Brazil the LGPD entered into force in September 2020, imposing a regulation similar to the GDPR. We consequently observe an average CMP adoption rate in South America of 26\% by the end of 2023.

In short, we clearly observe that the entrace into force and important changes to the regulations stimulate the need for valid technical solutions, creating a new market need that CMPs satisfy.

\subsection{The CMP market ecosystem}

\begin{figure}
    \centering
    \includegraphics[width=0.5\columnwidth]{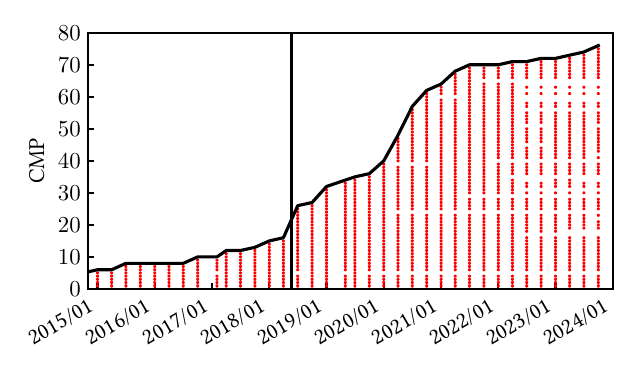}
    \caption{Presence of the considered CMPs in the observation period. \HTTPLong Dataset of 11,819 websites that are present during the whole period, in GDPR-regulated countries.}
    \label{fig:httparchive_cmp_time}
\end{figure}

We now study this new CMP market, discussing the pace at which they are born and how their market share evolves over time. We focus on the GDPR-based websites present in the \HTTPLong dataset. We first study when new CMPs enter this market over the years. In  Figure~\ref{fig:httparchive_cmp_time}, the $x$-axis represents time;  On the $y$-axis, we sort the CMPs by the time of their first appearance. A dot in the figure indicates that the i-th CMP was found on at least 10 different websites in that month. In fact, out of 84 CMPs present in our list, 76 have actually been adopted by more than 10 websites (we expect some to be active uniquely in countries other than GDPR). Figure~\ref{fig:httparchive_cmp_time} shows that, before the GDPR, less than 20 CMPs were active on the market. The GDPR fostered the born of new solutions, and, by the end of 2019, 35 CMPs were active. In 2020 we observe a second increase in the number of CMPs, with the adoption of 30 new platforms, a symptom of a more vast, yet competitive market for privacy-management solutions. In 2021, 2022 and 2023 we testify only less than 10 newcomers, hinting the European market is rather saturated. Finally, we notice that a non-negligible fraction of CMP (18 out of 76) has apparently shut down or has been acquired by other companies (i.e., is no longer present as testified by the missing dots).

\begin{figure}
\centering
\begin{minipage}{.45\textwidth}
  \centering
\includegraphics[width=\linewidth]{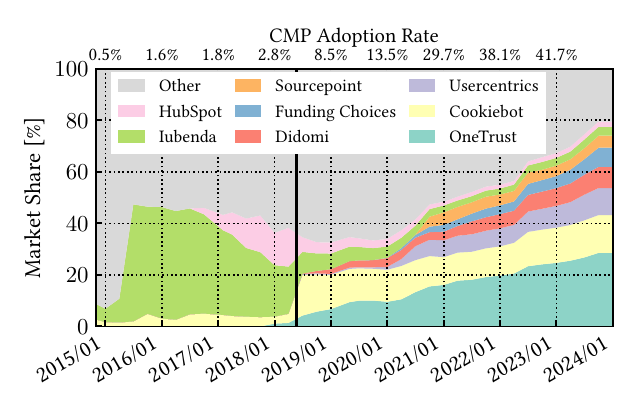}  \captionof{figure}{Market share of the top-8 CMPs. \HTTPLong Dataset of 11,819 websites that are present during the whole period in GDPR-regulated countries.}
  \label{fig:share-rel}
\end{minipage}%
\qquad
\begin{minipage}{.45\textwidth}
  \centering
  \includegraphics[width=\linewidth]{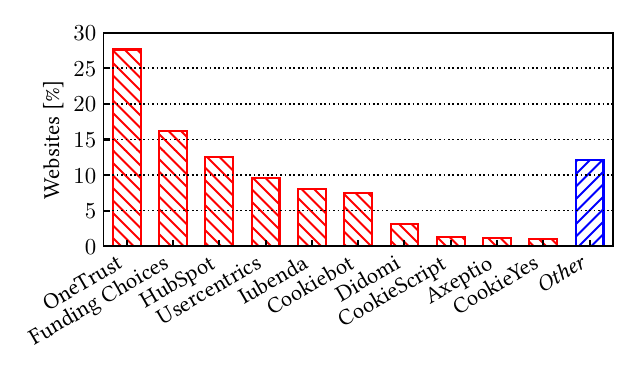}
  \captionof{figure}{Market share of the top-10 CMPs on the last snapshot. \HTTPLast Dataset with top-1M website worldwide.}
  \label{fig:cmp_last}
\end{minipage}
\end{figure}

We now focus on CMP market share. For each CMP, we compute its market share as the ratio between the number of websites adopting it and the number of websites using any CMP. In Figure~\ref{fig:share-rel} we report the market share evolution in the GDPR countries of the top-8 CMPs (as measured in the last time step of the dataset). The grey area represents the remaining CMPs. For reference, the top-$x$-axis reports the overall CMP adoption rate (recall it increases over time). Before the GDPR came into force, Iubenda dominated the market: it was the choice for more than 40\% of (mostly Italian) websites resorting to a CMP. The main competitor of Iubenda was Hubspot, even if with moderate penetration. After GDPR, the leading position of Iubenda is quickly eroded by new competitors: Cookiebot and OneTrust entered the market, the former reaching a market share of 16.1\% by the end of 2020. Since then Cookiebot maintained its market share of about 25-28\%. OneTrust, a company offering various marketing and analytics solutions, exhibits a significant growth that reaches 29\% at the end of 2023. This is also thanks to the acquisition of the CookieLaw and CookiePro competitors. The youngest notable player is Funding Choices, a CMP operated by Google, which appeared in 2020 and currently with a market share of around 7.4\%.

We complete the analysis by providing the most recent snapshot of the worldwide CMP market share. We rely on the top-ranked 1 million websites as of December 2023 in the \HTTPLast dataset, where 13.1\% of them adopts a CMP (see Table~\ref{tab:httparchive_global}). Figure~\ref{fig:cmp_last} details the market share of the top 10 CMPs worldwide. We find most of the CMP already presented in Figure~\ref{fig:share-rel}, even if with a different rank and share due to the different website bases. The leader is again OneTrust (27.6\% of market share). Funding Choices by Google comes second with 16.1\% of market share among the top-1M worldwide websites. Interestingly, its market share is only 7.4\% of European websites existing in all 9 years. This suggests Funding Choices CMP is especially adopted in non-GDPR countries. Conversely, CookieBot has a market share of 14.7\% on the \HTTPLong(GDPR) and only 7.3\% on \HTTPLast, suggesting it is more popular among  GDPR countries. 

To summarize, CMPs are products born as a result of European privacy regulations. Different countries show different patterns caused by their internal dynamics and specific implementations of the European directives and privacy regulations at large. While the market was initially dominated by a few players, nowadays different companies compete. The availability of longitudinal data as those provided by the HTTP Archive is fundamental for this type of analysis.

\section{User interaction}
\label{sec:user-interaction}

In this section, we focus on users' interaction patterns when facing a Privacy Banner. To this end, we use the proprietary dataset of the illow.io CMP described in Section~\ref{sec:illow-dataset-method}.

\begin{figure}
    \centering
    \includegraphics[width=0.5\columnwidth]{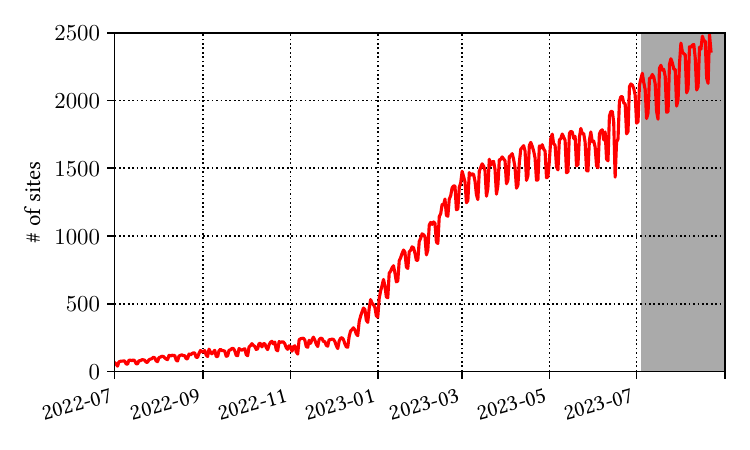}
    \caption{Evolution of the number of sites with at least 1 daily recorded interaction in the illow.io dataset.}
    \label{fig:illow-nsites-timeseries}
\end{figure}

In total, illow.io serves more than 6,000 websites all over the world, collecting hundreds of thousands of interactions daily. In detail, in Figure~\ref{fig:illow-nsites-timeseries}, we show the daily number of websites for which we observe at least one interaction. The weekly periodicity is due to Sundays which record the least amount of traffic. We first observe that the number of the CMP's customer websites grows over time, with a significant increase starting in December 2022 and a second one in June 2023. Since then, more than 2,000 websites have been active daily. Not shown for brevity, the number of daily interactions grows proportionally, surpassing 180,000 daily interactions in the last period.

\subsection{Longitudinal analysis}

\begin{figure}
    \centering
    \begin{subfigure}[b]{\columnwidth}
        \centering
        \includegraphics[width=0.8\textwidth]{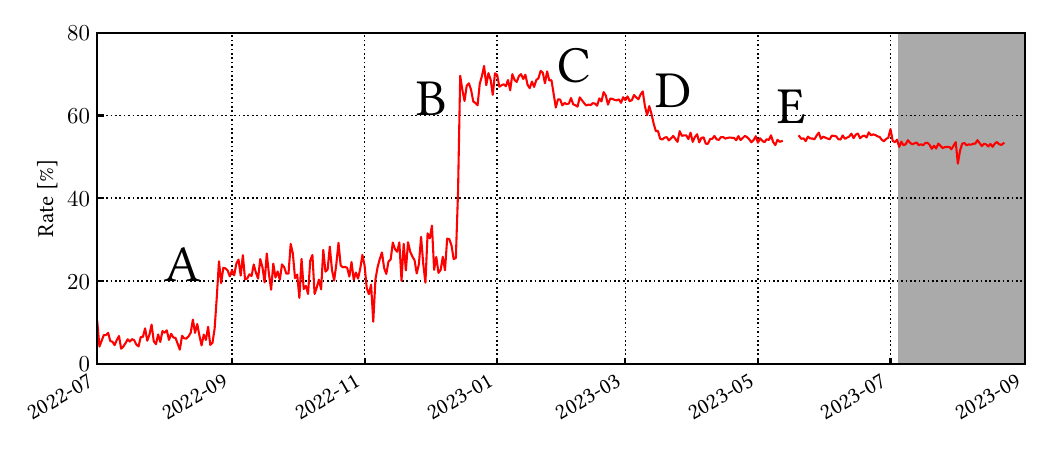}
        \caption{GDPR-regulated countries.}
        \label{fig:illow-gdpr-timeseries}
    \end{subfigure}
    \begin{subfigure}[b]{\columnwidth}
    \centering
        \includegraphics[width=0.8\textwidth]{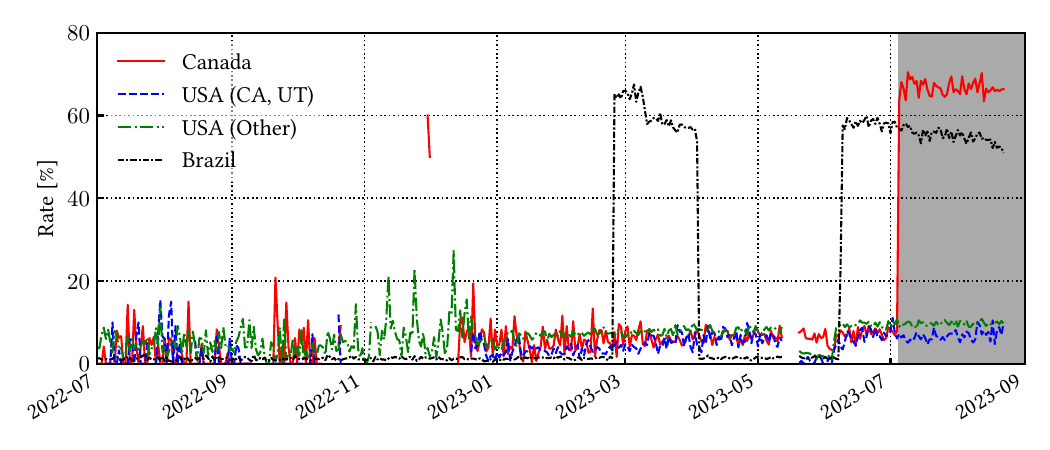}
        \caption{Other areas.}
        \label{fig:illow-rest-of-world-timeseries}
    \end{subfigure}
    \caption{The \rejected rate in a 14-months period by macro areas.}
    \label{fig:illow-timeseries}
\end{figure}

We first focus on the evolution of the fraction of \rejected interactions. Figure~\ref{fig:illow-timeseries} reports the daily average \rejected rate for users connecting from different areas of the world, separately for the GDPR area (Figure~\ref{fig:illow-gdpr-timeseries}) and other notable countries (Figure~\ref{fig:illow-rest-of-world-timeseries}). On the contrary to what we define in Section~\ref{sec:illow-method}, we remove the 50-websites constraint --- so that we have enough data for most of the days in out 14-months span. Figure~\ref{fig:illow-timeseries} thus shows the average among all the websites with at least ten daily interactions.

Starting from the GDPR area, observe how the curve follows a complicated shape. Each change is a consequence of changes in the banner design at illow.io. We highlight and describe the five main changes:

\begin{itemize}
    \item\textbf{{A -- Add \rejectbutton button}}: In August 2022, the CMP changed the banner shown to the GDPR users from the \textit{General} -- two-buttons banner -- to the \textit{GDPR-compliant} -- three-buttons banner. The new banner includes the \rejectbutton button, allowing users to refuse the usage of non-mandatory cookies with a simple one-click action. This addition is a consequence of the fine imposed by CNIL (the French data protection authority) on Google and Facebook in January 2022\footnote{\url{https://www.cnil.fr/sites/cnil/files/atoms/files/deliberation_of_the_restricted_committee_no._san-2021-023_of_31_december_2021_concerning_google_llc_and_google_ireland_limited.pdf}, Accessed on \today.}. This suddenly increases the \rejected rate to $\approx20\%$ (we already observed this behaviour in our previous work~\cite{jha2023i}). 
    \item\textbf{{B -- Add ``X'' button, activate blurring, top position}}: In December 2022, the CMP introduced the ``Close banner'' ``X'' button. Clicks on this button are equivalent to a \rejected choice. Following again the suggestions of European privacy-regulation bodies, The CMP also changed the \rejectbutton button to have the same colour as the \acceptbutton button.
    The banner is shown by blurring the background and placed in a central position on the page. This forces the users to interact with the banner to access the website content. Users quickly click on the ``X'' or \rejectbutton buttons. In a nutshell, 
    they are more interested in discarding the banner rather than making an informed decision. The reject rate tops 70\% (and 30\% explicitly accepting cookies).

    \item \textbf{{C -- Remove blurring; D -- central position}}: In January and March 2023, the CMP took some minor measures to reduce the \rejected rate: they disabled the blurring effect (C), and moved the banner in the bottom position by default (D). These changes allow the user to see the website content without getting rid of the banner, which causes a reduction in the \rejected rate. The \rejected rate drops to 55\%. We link this drop to users ignoring the banner (for which we have no records), rather than intentionally selecting to accept cookies.

    \item \textbf{{E --lack of data}}: In May 2023, the data collection was suspended for a short amount of time for technical reasons.
\end{itemize}

The same events, although differently combined and at different times, explain the shape that we observe in other regions in Figure~\ref{fig:illow-rest-of-world-timeseries}. Focus on Canada (red curve) and Brazil (black curve): The CMP introduced a GDPR-compliant banner for Brazilians in February 2023 with a configuration similar to A+B (i.e., with the \rejectbutton and ``X'' buttons). The presence of the intrusive banner suddenly makes the users close the banner by clicking on the ``X'' which is equivalent to selecting \rejected. In April 2023, the CMP defaults to C+D, reducing the \rejected rate by users. The same happens in Canada where users start interacting with the GPDR-compliant banner in July 2023.

For other countries, the CMP shows the General banner with neither the ``X'' nor the \rejectbutton button (see Figure~\ref{fig:banner-general}). The \rejected rate is far lower and barely reaches 10\%. In this case, a user who would like to disable the usage of profiling and tracking cookies has to perform multiple clicks. This burden makes the user prefer to either ignore the banner or quickly select the \acceptbutton button to dismiss the banner.

In a nutshell, users aim to quickly close the banner. Less than 10\% of them go through the burden of manually disabling the data collection when more than one click is needed. When the \rejectbutton button is present, users select it in 20\% of cases. When the ``X'' button is present (and equivalent to the \rejectbutton action), users just click on it to quickly close the privacy banner.

For completeness, the curves in Figure~\ref{fig:illow-timeseries} include the per-region average \rejected rate if at least 10 interactions are collected for the given samples (i.e., at least 10 interactions on a website by users from the given region). This causes the curves to be noisy and possibly lack some points, especially in the early months when the number of interactions is low -- see Figure~\ref{fig:illow-nsites-timeseries}.

\subsection{Stationary analysis}

\begin{figure}
    \centering
    \includegraphics[width=0.8\columnwidth]{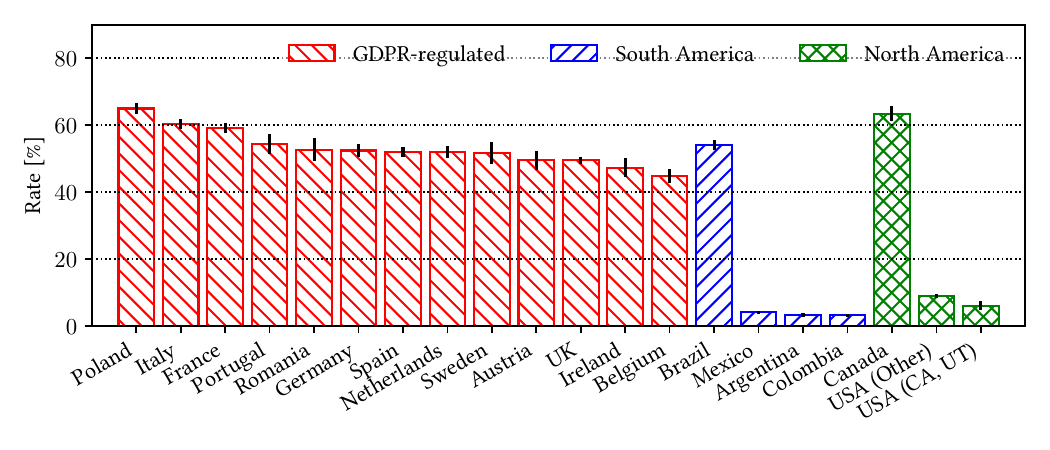}
    \caption{The \rejected rate per country. illow.io dataset from July 3\textsuperscript{rd}, 2023.}
    \label{fig:illow-country-barplot}
\end{figure}

We now focus on the last two months of illow.io data (from July 5\textsuperscript{th} to the end of August '23, highlighted in grey in previous plots). By doing so, we ensure that the CMP operation is not changing, allowing us to drill down the analysis.

In Figure~\ref{fig:illow-country-barplot} we break down the \rejected rate based on users' country. To only show the most represented countries, we show countries with at least 50 websites recording 50 visits in the observation period --- rather than only 10 visits. The main differences are ascribable to the different versions of the banner the users are shown. The \rejected rate varies between 45\% and 65\% for countries with a GDPR-compliant banner (which include Brazil and Canada). Conversely, countries with default banners do not exceed the 10\% \rejected rate. We attribute the differences among countries with the same banner to different cultural habits, as differences are consistent across all of our dataset dimensions. As an example, we compare the \rejected rate for Italy and UK in Appendix~\ref{sec:it-vs-uk}.

Peculiar is the case of California and Utah, where users face the CCPA-compliant banner (as in Figure~\ref{fig:banner-ccpa}). The \rejected rate for these two states (6.6\%) is only moderately lower than in the remaining states of the U.S. (8.8\%), where users are offered the General banner (as in Figure~\ref{fig:banner-general}. These numbers are interesting if we focus on how the two types of banners operate. The CCPA-compliant banner is based on the \emph{opt-out} mechanism -- i.e., if the user ignores or closes the banner, the consent is presumed. Conversely, with, the General banner, the user must explicitly \emph{opt in} to cookies, although opting in is faster and easier than opting out (which requires accessing the selection windows). The opt-in mechanism is considered more respectful of users' privacy, but our results show that, when the banner design encourages users to opt in, the practical effect is almost comparable to an opt-out banner.

\begin{figure}
    \centering
    \includegraphics[width=0.8\columnwidth]{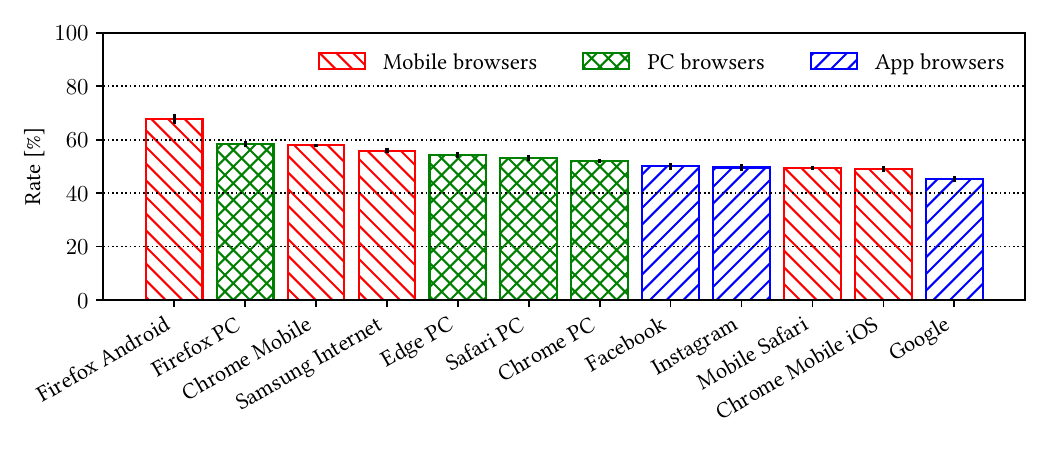}
    \caption{Average \rejected rate by browsers. illow.io dataset from July 1\textsuperscript{st}, 2023.}
    \label{fig:illow-bars-browsers}
\end{figure}

We now focus on other aspects and study how the user's browser and device impact the behaviour with privacy banners. In Figure~\ref{fig:illow-bars-browsers}, we compare the \rejected rate for users using different browsers to surf the Internet. We focus on users facing a GDPR-compliant banner, and we group the browsers into three categories for ease of visualization.
\begin{itemize}
    \item \textbf{Mobile browsers -- red pattern}: full browsers operating on a mobile device, such as a smartphone or a tablet. Among them, we include Chrome Mobile, Samsung Internet, and Firefox Mobile on Android devices, and the Mobile Safari and Chrome Mobile for iOS on iOS and iPadOS devices.
    \item \textbf{PC browsers -- green pattern}: full browsers operating on a laptop or desktop computer. We include Google Chrome, Safari, Edge, and Firefox.
    \item \textbf{App browsers -- blue pattern}: in-app browsers offered by popular applications on smartphones and tablets. These apps let the user browse websites in the app when they click a link. We observe a significant usage of Facebook, Instagram, and Google Search apps directly requesting and showing webpages to users.
\end{itemize}

Interestingly, Firefox stands first among both Android and PC browsers. Although we cannot prove any hypothesis on this outcome, we argue Firefox users are careful about their privacy and tend to refuse the use of cookies more often than other users. At the opposite edge, users navigating app browsers tend to refuse less. This behaviour might have multiple causes: for example, users inside a third-party app's environment might be less encouraged to make an informed choice about cookies, given the unusual surfing context they are in. Or the more occasional browsing makes them quickly return to the app without spending time on the webpage content.

\begin{figure}
    \centering
    \includegraphics[width=0.4\columnwidth]{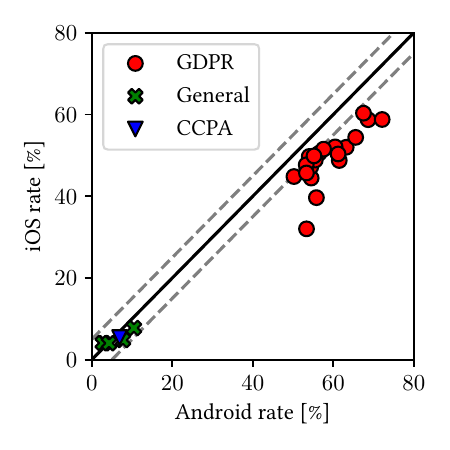}
    \caption{Android vs iOS: Average \rejected rate for users in each country, since July 3\textsuperscript{rd}, 2023. The dotted lines represent a $\pm5\%$ difference.}
    \label{fig:illow-scatter-android-ios}
\end{figure}

We now draw attention to mobile, comparing the behaviour of Android and iOS users, for which we find notable differences. In general, we find that Android users have high \rejected rates (see for example Firefox Android in Figure~\ref{fig:illow-bars-browsers}): this may be caused by the large relative banner area which covers a large fraction of the screen. Users quickly dismiss the banner by clicking on the ``X'' more frequently on a mobile than on a desktop browser (we will investigate this later). However, for iOS users, the picture is different -- for instance, Mobile Safari and Chrome Mobile iOS stand in the last positions in Figure~\ref{fig:illow-bars-browsers}. We explore the remarkable differences between Android and iOS users more in-depth in Figure~\ref{fig:illow-scatter-android-ios}.
For every country, we plot the \rejected rate for users using Android versus iOS operating system. We consider countries with at least 10 websites with 50 interactions each.
The diagonal dashed lines represent a $\pm5\%$ difference. Interestingly enough, in every country we observe iOS users present a significantly smaller \rejected rate than Android users. Our data cannot offer us any insight into the reason for this difference. We hypothesize this could be due to iOS users being more confident about their privacy than their Android counterparts because of the feeling of privacy-friendly solutions offered by the Apple ecosystem.

\begin{figure}
    \centering
    \includegraphics[width=0.8\columnwidth]{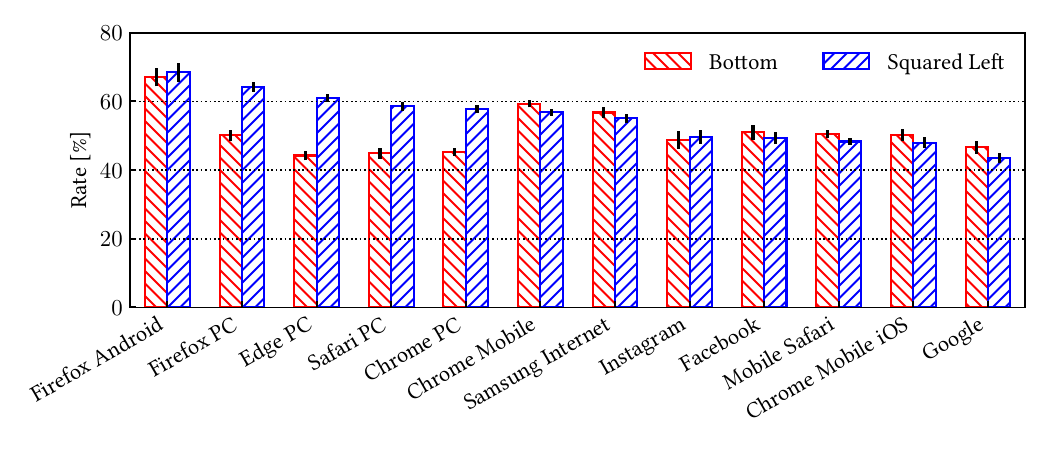}
    \caption{Average \rejected rate by browsers, divided by banner position. illow.io dataset since July 3\textsuperscript{rd}, 2023.}
    \label{fig:illow-bars-position}
\end{figure}

At last, we differentiate the outcome with respect to the Privacy Banner position on the screen. We consider the two most popular banner options: \textit{Bottom}, i.e., the banner appears on the lower part of the screen - covering a portion of the bottom area with an overlay window; \textit{Squared Left}, i.e., the banner appears as a pop-up in the bottom-left corner of the screen (see Figure~\ref{fig:banners}), resulting more visible. In Figure~\ref{fig:illow-bars-position}, we observe that the position of the banner makes a significant difference only on Desktop browsers. In fact, the banners overlayed on the bottom of the page register a lower \rejected rate than the Squared Left pop-up style banner. This hints at users being annoyed by the pop-up banner window, nudging more people to dismiss it via the ``X'' button to freely access the page. In Mobile and App browsers on smartphones and tablets, there is no significant difference in the \rejected rate between the two types of banners. This is because there is practically no difference between the layout of the Bottom and of the Squared Left banners - both covering a large fraction of the mobile device screen.

\subsection{Other findings}

Here we report some additional findings regarding users' interaction with the illow.io Privacy Banner.
\begin{itemize}
    \item{Access to Cookie Policy and the Privacy Policy}: from August 2022 to April 2023, the illow.io dataset reports clicks on the Cookie Policy and the Privacy Policy links offered by the banner. Only 0.22\% of users explore this possibility and click on at least one of the two policies. It stands clear that common website visitors are not interested in spending time reading long and detailed policy text.
    \item{Fine-grained consent}: few users choose to fine-tune the categories of cookies they are willing to accept. In the whole illow.io dataset, we only count 8,297 interactions which offered a partial acceptance over almost 25 million interactions. The statistics category is the most accepted one (62.77 $\pm$ 1.04\%), followed by preferences (52.95 $\pm$ 1.07\%). The marketing category (arguably the one most associated with the feeling of privacy lack) is the least accepted one (19.63 $\pm$ 0.85\%). All confidence intervals are computed with a 95\% probability on a per-website grouping.
\end{itemize}
\section{Conclusion}
\label{sec:conclusion}

In this paper we discussed different aspects related to Consent Management Platforms, considering both their penetration into the web ecosystem and the way users interact with them.

Thanks to the long-lasting effort of the HTTP Archive, we showed the significant growth of CMP adoption on the web to ease the management of different and evolving privacy regulations. We witnessed that the CMP adoption significantly varies according to their region and when the privacy regulations entered into force.
Thanks to the GDPR, Europe leads the CMP adoption, with other regions of the world growing quickly when regulation enters into force, generating new opportunities for the CMP market.

Our results also suggest that some websites and CMPs still allow the browser to contact tracking systems and let them install persistent cookies before the user accepts their use. However, the limitations of the HTTP Archive call for further investigations to verify whether this corresponds to an actual violation of privacy legislation. At large, this shows the need to extend the availability of archiving platforms like HTTP Archive due to the intertwining of the regulations, user location, device type, language, etc. In fact, the customisation of content websites serve to users includes the way CMPs show the privacy banners and their actions.

Considering users' interactions with CMPs, our results showed that, when offered a viable option to refuse data collection, users tend to do so. Even more clearly, we showed that a large percentage of users tend to dismiss the Privacy Banner without actively interacting with it, but simply clicking on the close button if present, without making any explicit decision about their privacy. Such behaviour challenges the concept that offering users fine-grained options about their privacy empowers them to make the best decision. For instance, when bugged with more intrusive banners, the reject rate increases inflated by the need to close the banner. Conversely, when refusing data collection is cumbersome and requires more than one click, users tend not to take the opportunity. Interestingly, with an opt-out banner, where, by default, consent is presumed, the rejection rate is similar.

Finally, our results showed that users operating on mobile devices are apparently the most keen to refuse the use of data collection -- in fact they just to counterbalance the intrusiveness of the banner on mobile device screens. Curiously, iOS users tend to refuse at a lower rate than Android ones. Although we formulate some hypotheses about it, this observation poses the ground for interesting future work.

\begin{acks}
This work was partially supported by project SERICS (PE00000014) under the MUR National Recovery and Resilience Plan funded by the European Union - NextGenerationEU and the project "National Center for HPC, Big Data and Quantum Computing", CN00000013 (Bando M42C – Investimento 1.4 – Avviso Centri Nazionali” – D.D. n. 3138 of 16.12.2021, funded with MUR Decree n. 1031 of 17.06.2022). 
\end{acks}

\bibliographystyle{ACM-Reference-Format}
\bibliography{biblio}

\appendix

\section{Dataset structures}

In this Appendix, we provide the exact structure of our two datasets. Both datasets are in tabular format, and, in the following paragraphs, we report the columns relevant to our analysis (we omit other columns, useless for our purposes).

\subsection{HTTP Archive}

We use the public data offered by the HTTP-Archive to build the \HTTPLong and \HTTPLast datasets. The pre-processed data and the code to obtain the figures of this paper are available online~\cite{opendata}. In particular, we downloaded via Google Cloud Storage the tables called \emph{Summary Pages} and \emph{Summary Requests}, containing details on the visited pages and all issued HTTP requests, respectively. We downloaded the tables for the 9 years of interest. Relevant columns are:
\begin{itemize}
    \item \textbf{Summary Pages}:
    \begin{itemize}
        \item Numerical identifier of the visit
        \item URL of the visited website
        \item Time of the visit
    \end{itemize}
    \item \textbf{Summary Requests}:
    \begin{itemize}
        \item Identifier of the corresponding visit
        \item URL requested by the browser in the HTTP request
        \item Presence of the \texttt{Set-Cookie} header. The dataset does not include the content of the Cookie.
    \end{itemize}
\end{itemize}

\subsection{illow.io CMP dataset}

This dataset is contained in a single data table, where each entry corresponds to an \emph{interaction}. The columns relevant to our analysis are:
\begin{itemize}
    \item Time of the interaction
    \item Anonymous identifier of the interaction
    \item Anonymous identifier of the website
    \item Client's $/24$ IPv4 subnet
    \item Client's \texttt{User-Agent} as set by the browser in the HTTP requests
    \item Consents Provided among the following Cookie categories:
    \begin{itemize}
        \item Necessary (always present)
        \item Statistical
        \item Preferential
        \item Marketing
    \end{itemize}
\end{itemize}

\section{Italy vs UK}
\label{sec:it-vs-uk}

Here, we compare the \rejected rate of Italy and the UK to show that consistent differences exist across all of the dimensions of our dataset. We presume that the differences in rates between the two countries (as shown in Figure~\ref{fig:illow-country-barplot}) only depend on users' cultural habits.

\begin{figure}
    \centering
    \includegraphics[width=.8\textwidth]{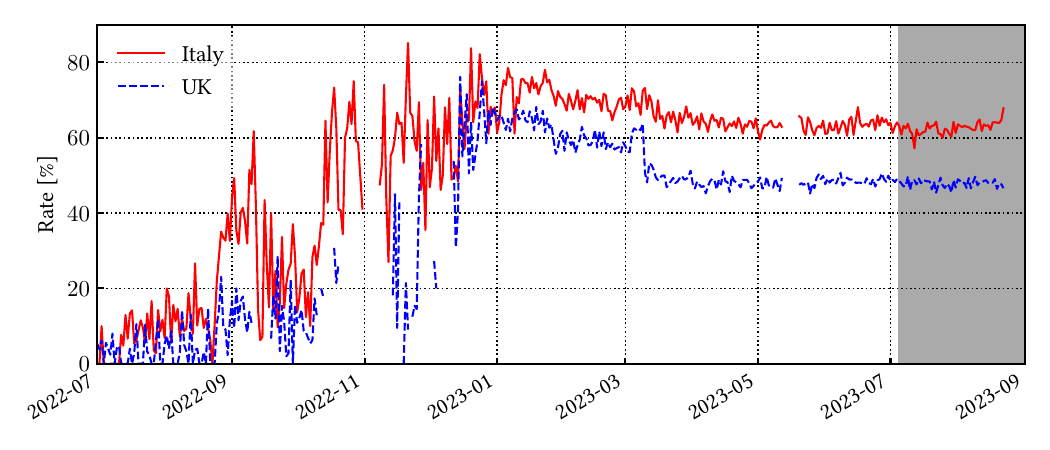}
    \caption{The \rejected rate of Italy and UK.}
    \label{fig:illow-it-vs-uk-timeseries}
\end{figure}

In Figure~\ref{fig:illow-it-vs-uk-timeseries}, we compare the \rejected rate by Italian and British users. The distance between the two countries has been sizeable and stable since the beginning of 2023. We can thus exclude that any variation in the CMP operation has biased the measurement.

\begin{figure}
    \centering
    \includegraphics[width=.8\textwidth]{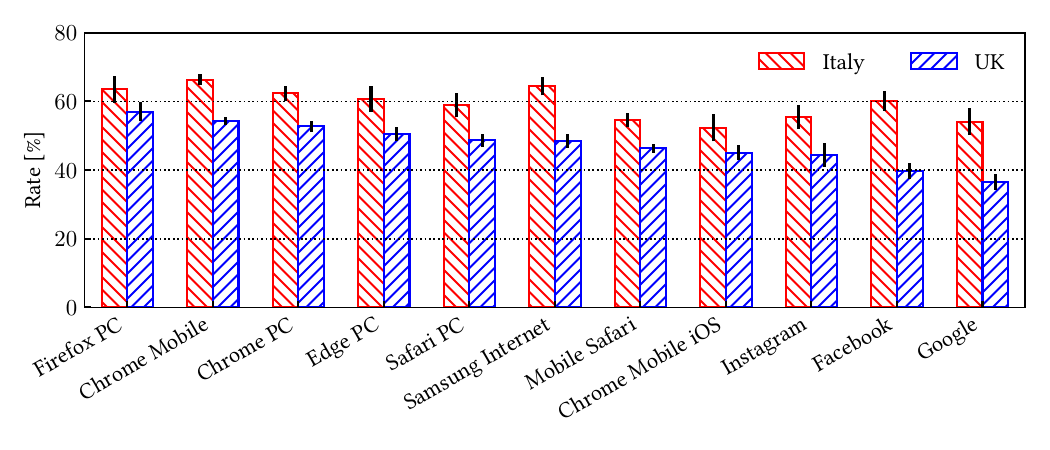}
    \caption{The \rejected rate of Italy and UK.}
    \label{fig:illow-it-vs-uk-browser}
\end{figure}

In Figure~\ref{fig:illow-it-vs-uk-browser}, we partition the dataset across the most popular browsers. The \rejected rate for Italy is always higher than for the UK, regardless of the considered browser (although for some of them, the distance is minimal -- e.g., Firefox PC). We can also rule out that the difference between the two countries is due to the different browser popularity. With the information at our disposal, we thus assume the difference is attributable to different cultural habits among the populations.

\end{document}